\documentclass[prb,twocolumn,showpacs,amsfonts,amssymb,floats,superscriptaddress,aps]{revtex4}
\usepackage[final,dvips]{epsfig}

\begin{document}

\title[]
{Conductance Through a Redox System in the Coulomb Blockade Regime: Many-Particle Effects and Influence of Electronic Correlations}

\author{
Sabine Tornow}
\affiliation{\mbox{Institut f\"ur Mathematische Physik, TU Braunschweig, 38106 Braunschweig, Germany}}
\author{
 Gertrud Zwicknagl}
\affiliation{\mbox{Institut f\"ur Mathematische Physik, TU Braunschweig, 38106 Braunschweig, Germany}}
\date{\today}

\begin{abstract}
We investigate the transport characteristics of a redox system weakly coupled to leads in the Coulomb blockade regime. The redox system comprises a donor and acceptor separated by an insulating bridge in a solution. It is modeled by a two-site extended Hubbard model which includes on-site and inter-site Coulomb interactions and the coupling to a bosonic bath. The current voltage characteristics is calculated at high temperatures using a rate equation approach. For high voltages exceeding the Coulomb repulsion at the donor site the calculated transport characteristics exhibit pronounced deviations from the behavior expected from single-electron transport. Depending on the relative sizes of the effective on-site and inter-site Coulomb interactions on one side and the reorganization energy on the other side we find negative differential resistance or current enhancement.
\end{abstract}
\maketitle

\begin{figure}
\hspace{-1cm}
\vspace{1cm}
\includegraphics[width=60mm]{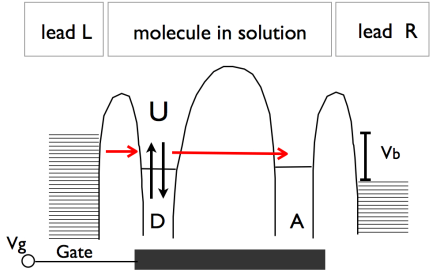}
\caption{Schematic view of the redox system with donor (D) and acceptor (A) coupled to the leads L and R. The electronic degrees of freedom of the DA system are coupled to the environment comprising internal vibrations and the solvent dynamics. The current is calculated as a function of the bias voltage $V_b$ and gate voltage $V_g$. }
\label{Schema}
\end{figure}
Electron transfer is fundamental in chemistry and biology. 
In a polar environment the redox process involves strong coupling to the underlying nuclear motion and is usually dominated by the nuclear reorganization that accompanies the charge rearrangement. Such a redox system contacted to leads can serve as a component of molecular electronic devices \cite{RatnerAviram}. Recent experiments \cite{Tao} showed that the electron transport in redox systems depends strongly on the environment (solvent) and can be described by a thermally activated electron transfer process mediated by thermal fluctuations of the environment.

\begin{figure}
\hspace{-1cm}
\vspace{-2cm}
\includegraphics[width=184mm]{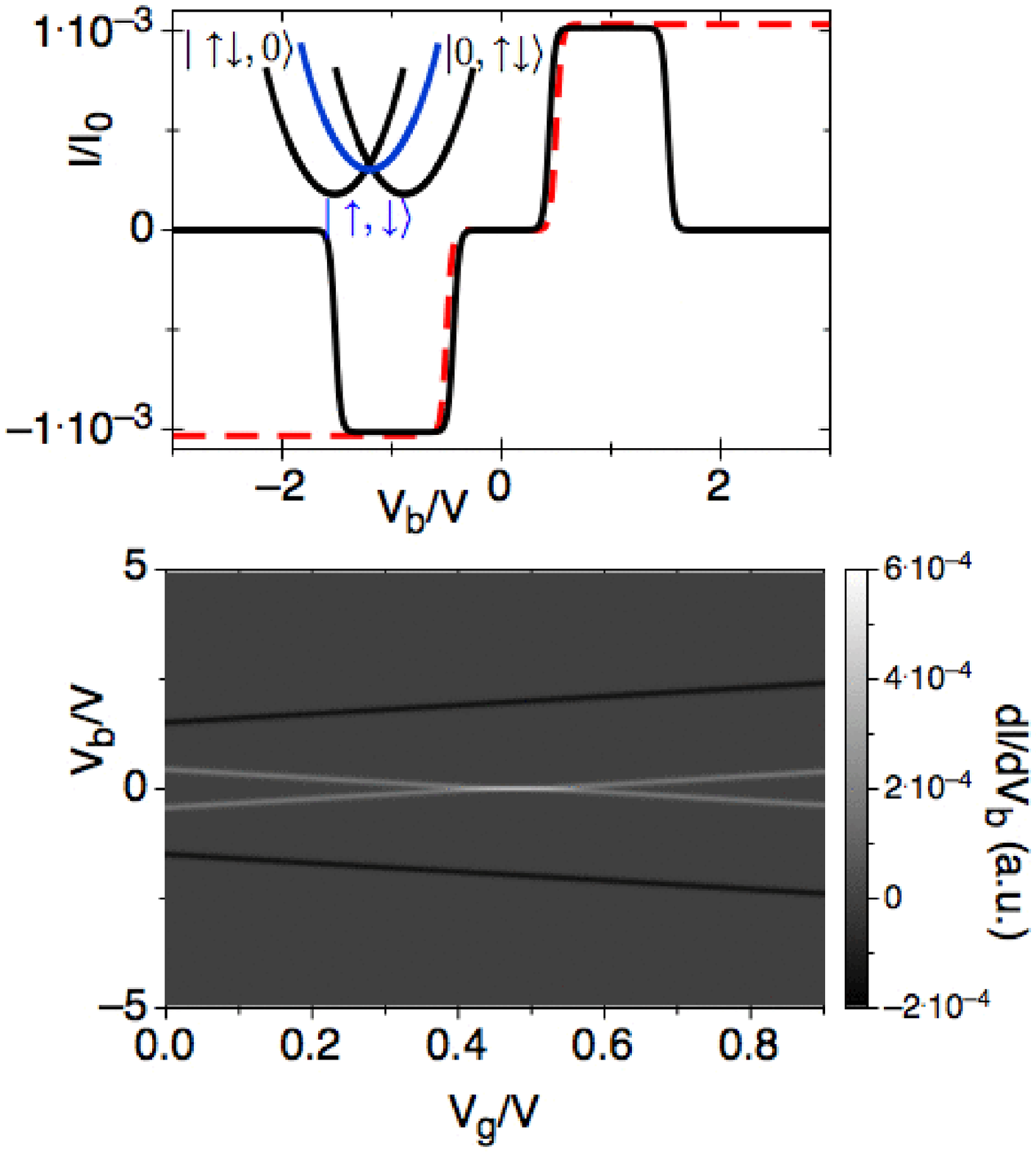}
\vspace{-2cm}
\caption{(color online) Current $I$ as a function of $V_b$ for $U_{\rm eff}=1eV$ and $U_{DA}=2 eV$ (upper panel, black solid line) in comparison to the case of independent electrons (upper panel, red dashed line) as well as a contour plot of $dI / dV$ vs. $V_g$ and $V_b$ for correlated electrons (lower panel). The other parameters see in the text. Inset: Potential surfaces for the states: $|\uparrow \downarrow,0 \rangle$, $|\uparrow,\downarrow \rangle$ and $|0, \uparrow \downarrow \rangle$.  }
\label{Bild1}
\end{figure}

\begin{figure}
\vspace{-1cm}
\hspace{-2cm}
\includegraphics[width=184mm]{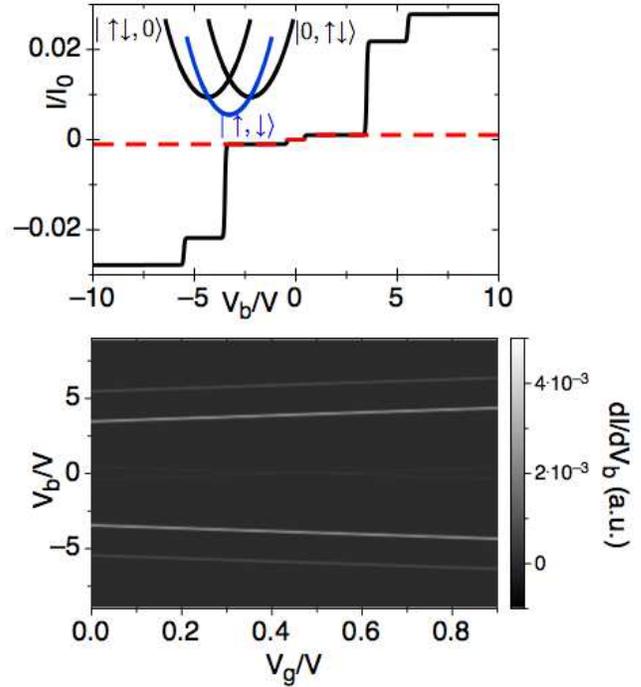}
\vspace{-2cm}
\caption{(color online) Current $I$ as a function of $V_b$ for $\tilde{U}_{\rm eff}=3 eV$ and $V=1 eV$ (upper panel, black solid line) in comparison to the case of independent electrons (upper panel, red dashed line)  as well as a contour plot of $dI / dV$ vs. $V_g$ and $V_b$ for correlated electrons (lower panel). The other parameters see in the text. Inset: Potential surfaces for the states: $|\uparrow,\downarrow,0 \rangle$, $|\uparrow,\downarrow \rangle$ and $|0, \uparrow \downarrow \rangle$. Note that in this case the transfer from $|\uparrow\downarrow,0 \rangle$ to $|\uparrow,\downarrow \rangle$ is in the activation-less regime.}
\label{Bild2}
\end{figure}

Although the standard description focuses on single-electron transfer \cite{Transfer} multiple electron transfer is the dominant process in some chemical systems \cite{Evans,Two,May}.
Many-particle states have to be taken into account and the rate between these states depends strongly on the Coulomb interaction \cite{Tornow}.
Contacting such a redox-system to leads and applying a voltage larger than the effective on-site Coulomb repulsion, e.g., on the donor, it can be occupied by multiple electrons. Then the single-electron picture breaks down and the transport is affected by electron correlations \cite{Ghosh,Parida}. In the present paper, we study the current through such a redox sytem taking the solvent explicitly into account for the first time. The central results are displayed in Fig.˜\ref{Bild1} and Fig.˜\ref{Bild2} where negative differential resistance (NDR) or current enhancement is be found, respectively.
These effects occur at voltages when multiple electrons populate the DA system.


We consider a two-site model where the redox system comprises donor (D) and acceptor (A) which are separated by an insulating bridge (see Fig. \ref{Schema}). The system is embedded in a solvent and weakly coupled to metallic electrodes. The Hamiltonian is given by $H=H_{\rm DA}+H_{\rm DA-C}+H_{\rm C}$ where the DA system is described by

\begin{equation}
H_{\rm DA}=\sum_{\sigma; i=D,A} \epsilon_i n_{i \sigma} -\Delta
\sum_{\sigma} \left( d_{D \sigma}^{\dag} d_{A \sigma}+d_{A \sigma}^{\dag} d_{D \sigma} \right) 
\nonumber
\end{equation}
\begin{equation}
+U \sum_{\sigma; i =D,A} n_{i \sigma } n_{i -\sigma}  + U_{DA}
\sum_{\sigma,\sigma'} n_{D \sigma} n_{A \sigma'} 
\end{equation}
\begin{equation}
+\sum_{\sigma}\left(n_{D,\sigma}-n_{A,\sigma}\right) \sum_{n}
         \lambda_n \left(  b_{n}^{\dagger} + b_{n}
         \right)+\sum_{n} \omega_{n} b_{n}^{\dagger} b_{n} .
\nonumber
\label{coupl_bos}
\end{equation}
As written $c_{i\sigma}$ and $c^{\dagger}_{i\sigma}$ denote annihilation
and creation operators for fermions with spin $\sigma$ in an orbital
localized at site $i$ and $n_{i,\sigma} = c^{\dagger}_{i\sigma} c_{i\sigma}$ is the number-operator. The first four terms correspond to an extended 
two-site Hubbard model, with on-site energies $\epsilon_i$, hopping 
matrix element $\Delta$, on-site Coulomb repulsion $U$, and inter-site Coulomb repulsion $U_{DA}$. 
The two last parts describe the electron-boson coupling (in the standard polaron form) and the free bosonic bath with boson creation and annihilation operators $b_n^{\dagger}$ and $b_n$, respectively. Furthermore, the coupling between leads and DA system is given by
\begin{equation}
H_{\rm DA-C} =\sum_{k,\sigma} \left(V_k c_{L,k,\sigma}^{\dagger} d_{D,\sigma} +V_{k} c_{R,k,\sigma}^{\dagger} d_{A,\sigma}+h.c.)\right)
\end{equation}
while the dynamics of the free fermions in the leads is accounted for by
$H_{\rm C}=\sum_{k,\sigma;i=L,R} \epsilon_k n_{i,k,\sigma} $.
In the following, we focus on the Coulomb blockade regime with weak coupling between leads and DA system which is strongly coupled to the bosonic bath. The electrons tunnel sequentially on and off as well as migrate via thermal activation through the DA system. At high temperatures, exceeding the activation energy, Marcus approximation is valid where only one bosonic mode with frequency $\omega_0$ is considered \cite{Tornow}. We obtain an effective model where the parameters are renormalized according to:  $\tilde{\epsilon}_{D,A} = \epsilon_{D,A}-E_{\lambda}$, $U_{\rm eff}=U-4E_{\lambda}$ (with the reorganization energy $E_{\lambda}=\lambda^2/\omega_0$, $\Delta \rightarrow \tilde{\Delta}$ and $V_k \rightarrow \tilde{V}_k$. 

The DA-system has 16 basis-states $|D,A\rangle $ describing the occupation on D (A),
with different number $n=n_D+n_A$ of electrons where $n_D (n_A)$ is the number of electrons on D (A): $|\uparrow, 0\rangle $,$|\downarrow, 0\rangle $, $|0,\uparrow\rangle$, $|0,\downarrow\rangle$,$|\uparrow,\uparrow \rangle$, $|\downarrow,\downarrow \rangle$, $|\uparrow \downarrow ,0\rangle$,$|0,\uparrow \downarrow \rangle$, $| \uparrow \downarrow,\uparrow \rangle $, $| \uparrow \downarrow,\downarrow \rangle $, $| \uparrow ,\uparrow \downarrow\rangle$,$| \downarrow ,\uparrow \downarrow\rangle$,$|\uparrow \downarrow, \uparrow \downarrow\rangle$ and $|0,0\rangle$.

Assuming fast dephasing and incoherent transfer the population dynamics can be described by kinetic equations:
\begin{eqnarray}
\dot{P}_s=\sum_{r} k_{r \rightarrow s} P_r-\sum_{r} k_{s \rightarrow r} P_s,
\end{eqnarray}
where $P_s$ is the probability that the system is in state $s$ and
$k_{r \rightarrow s}$ is the 
rate of the transition from state $r$ to state $s$ which is specified in the following. If the  occupation $n_D$ is changed by one at fixed $n_A$ the rate is
\begin{eqnarray}
k_{s \rightarrow r,L}^{(n_D,n_A)\rightarrow (n_D+1,n_A )}=\tilde{\Gamma} f(\Delta E_{sr}-\mu_L)  |\langle r| d_D^{\dagger} | s \rangle |^2 .
\end{eqnarray}
For the reverse process we find
\begin{eqnarray}
k_{r \rightarrow s,L}^{(n_D,n_A)\rightarrow (n_D-1,n_A )}=\tilde{\Gamma} (1-f(\Delta E_{sr}-\mu_L )) \left|\langle 
s| d_D | r \rangle \right|^2 \ .
\end{eqnarray}
The hybridization strength $\tilde{\Gamma}=2\pi \sum_k |\tilde{V}_k|^2 \delta(\omega -\epsilon_k)$ is assumed to be energy independent and equal for R and L, $f$ is the Fermi distribution and $\rho$ is the density of states of the leads.  We assume $k_B T \gg \hbar\tilde{\Gamma}$. Similar rates are obtained if $n_A$ is changed and $n_D$ is constant by exchanging $L$ with $R$ and $D$ with $A$. The chemical potential of the lead L and R is defined
with $\mu_{L,R}=\epsilon_F \pm eV_b$, where $V_b$ is the bias voltage. A finite gate voltage leads to $\tilde{\epsilon}_D \rightarrow \tilde{\epsilon}_D+V_g$ and $\tilde{\epsilon}_A  \rightarrow \tilde{\epsilon}_A+V_g$.

The transfer rates, at conserved particle number $n$, is given by the Marcus theory in the nonadiabatic limit ($\tilde{\Delta} \ll \omega_0$)

\begin{equation}
k_{s \rightarrow r}^{(n_D,n_A)\rightarrow (n_D\pm 1,n_A \mp 1)}=2\pi \tilde{\Delta}^2 F(\Delta E_{sr}),
\end{equation}
\begin{equation}
k_{s \rightarrow r}^{(n_D,n_A)\rightarrow (n_D\pm2,n_A \mp 2)}=2\pi \frac{\tilde{\Delta}^4}{|U_{\rm eff}-U_{DA}|^2}  F(\Delta E_{sr}),
\end{equation}
where the Franck-Condon integral
\begin{eqnarray}
F(\Delta E_{sr})=\frac{1}{\sqrt{4 \pi E_{\lambda}^{s \leftrightarrow
r} k_B T}} e^{- \frac{(\Delta E_{sr}-E_{\lambda}^{s
\leftrightarrow r})^2}{4 E_{\lambda}^{s \leftrightarrow r} k_B
T}}
\end{eqnarray}
depends on the energy difference $\Delta E_{sr}$ between state $s$
and $r$ and the corresponding
reorganization energy  $E_{\lambda}^{s \leftrightarrow r}$, e.g., $E_{\lambda}^{|\uparrow,0\rangle \leftrightarrow |0,\uparrow\rangle}=E_{\lambda}$,  $E_{\lambda}^{|\uparrow \downarrow,0\rangle \leftrightarrow |\downarrow,\uparrow\rangle}=4 E_{\lambda}$ and $E_{\lambda}^{|\uparrow \downarrow,0\rangle \leftrightarrow | 0,\downarrow \uparrow\rangle}=16 E_{\lambda}$

We calculate the stationary current ($I$) as a function of bias voltage $V_b$ and gate voltage $V_g$
as
\begin{equation}
I /I_0 = \sum_{r,s} (k_{s \rightarrow r,L}^{n_D \rightarrow n_D+1} P_{r}- k_{r \rightarrow s,L}^{n_D \rightarrow n_D-1} P_{s})
\label{Strom}
\end{equation}
by assuming $\dot{P}_s=0$, $\sum_{s} P_{s}=1$ and 
for the parameters $1/(k_BT)=40eV$, $\tilde{\Gamma}=0.05eV$, $\tilde{\Delta}=0.1eV$, $E_{\lambda}=0.5eV$. The current depends on the rate of electron transfer from D to A. Neglecting electron correlations two electrons on D are transferred each with a single-particle Marcus rate. However, electron correlations have to be taken into account (even for $U=U_{DA}=0$) for a correct description, e.g., the rate for $|\uparrow \downarrow, 0 \rangle \rightarrow |\uparrow,\downarrow \rangle$ depends exponentially on $U_{\rm eff}-U_{DA}-4 E_{\lambda}$.
In Fig.˜\ref{Bild1} and Fig.˜\ref{Bild2} (upper panel) we display $I$ for $U_{\rm eff}=1eV, U_{DA}=2eV$ and $U_{\rm eff}=3 eV, U_{DA}=1 eV$, respectively, in comparison to the current where electron correlations are neglected (red dotted line) for $V_g=0$.
 For $eV_b < U_{\rm eff}$ both curves are identical but at $eV_b>U_{\rm eff}$ two electrons occupy D. The electrons are transferred via a transition from state $|\uparrow \downarrow,0 \rangle$ to $|\uparrow,\downarrow \rangle$, $|\downarrow,\uparrow \rangle$ or $|0, \downarrow\uparrow \rangle$ . The potential surfaces for these many-particle states are displayed in the inset. In Fig.˜\ref{Bild1} the transfer is in the normal regime and in Fig. \ref{Bild2} in the activation-less regime since $U_{\rm eff}-V=4 E_{\lambda}$. In the latter case the current is enhanced for $eV_b>U_{\rm eff}=3eV$ by a factor of 10000 in comparison to the uncorrelated case. For $(U_{\rm eff}-V-4 E_{\lambda})^2>4 E_{\lambda} k_BT$ the transfer is in the normal or inverted region and negative differential resistance (NDR) is present as seen in Fig.˜\ref{Bild1} (lower panel). 
For our parameters ($U_{\rm eff}=1eV, U_{DA}=2eV$) the current is surpressed by a factor of 20.
Furthermore, we examined a small dependency of the on-site energies on $V_b$ and found that the results do not change qualitatively.

In summary, we considered a solvated redox system weakly coupled to leads at high temperatures which is able to accept more than one electron. If the bias voltage is larger than the effective Coulomb repulsion, the donor can be occupied by two electrons and the transfer rates are dominated by correlation effects. We found that the current strongly depends on the difference between the on-site and inter-site Coulomb repulsion as well as the reorganization energy. The current is enhanced in the activation-less regime when the difference is zero, otherwise NDR behavior is obtained.
\section*{Acknowledgment}
This research was supported by the State of Lower- 
Saxony and the VolkswagenFoundation. We are grateful to A. Nitzan for helpful discussions.


\end{document}